\documentclass[letterpaper,twocolumn,10pt]{article}
\PassOptionsToPackage{table,dvipsnames}{xcolor}
\usepackage{styles/usenix-2020-09}
\usepackage{xcolor}
\usepackage{graphicx}
\usepackage{booktabs}
\usepackage{tabularx}
\usepackage{array}
\usepackage{makecell}
\usepackage{multirow}
\usepackage{amsmath}
\usepackage{amssymb}
\usepackage[ruled,vlined]{algorithm2e}
\usepackage{caption}
\usepackage[labelformat=simple]{subcaption}
\usepackage{xurl}
\usepackage{tikz}
\usetikzlibrary{arrows.meta,positioning}

\usepackage{enumitem}
\setlist{topsep=0pt, itemsep=0pt, parsep=0pt, leftmargin=1em}
\usepackage[compact]{titlesec}
\titlespacing*{\paragraph}{0pt}{2pt}{6pt}
\usepackage{cleveref}

\makeatletter
\renewcommand{\sectionautorefname}{\S\@gobble}
\renewcommand{\subsectionautorefname}{\S\@gobble}
\renewcommand{\subsubsectionautorefname}{\S\@gobble}
\makeatother

\hypersetup{
  colorlinks=true,
  linkcolor=blue!90!red,
  citecolor={red!60!black},
  urlcolor={blue!60!black}
}

\usepackage{xspace}
\usepackage{pifont}

\newcommand{\sys}{\textsc{Unimatch}\xspace}

\newcommand{\para}[1]{\vspace{1mm}\noindent\textbf{#1}\xspace}

\newcommand{\step}[1]{\ding{\numexpr181+#1\relax}}

\definecolor{fgred}{HTML}{B24C4C}
\definecolor{bgblue}{HTML}{3B6FB6}

\makeatletter
\@ifundefined{proposition}{}{}
\makeatother

\newlength{\titleboxheight}
\makeatletter
\def\@maketitle{\newpage
 \vbox to \titleboxheight{
 \vspace*{\fill}
 \vskip 2em
 \begin{center}{\Large\bf \@title \par}\vskip 0.375in minus 0.300in
  {\large\it
   \lineskip .5em
   \begin{tabular}[t]{c}\@author
   \end{tabular}\par}\end{center}\par
 \vspace*{\fill}
 }
}
\makeatother

\begin{document}

\date{}

\title{Different Parallelism, Same Bits:\\Bitwise-Consistent LLM Reinforcement Learning with \sys}
\author{{\rm Honglin Zhu$^{1,*}$\quad Tong Xin$^{2,*,\dagger}$\quad Jingsheng Yan$^{2,\dagger}$\quad
   Yingming Mao$^{3,4,\dagger}$\quad Jiaping Cao$^{1}$}\\
  {\rm Yongguang Lin$^{1}$\quad Mingkai Dong$^{2,\ddagger}$\quad Wenxi Zhu$^{1,\ddagger}$\quad
   Haibo Chen$^{2,\ddagger}$}\\[0.5em]
  $^{1}$Tencent\qquad
  $^{2}$Institute of Parallel and Distributed Systems, Shanghai Jiao Tong University\\
  $^{3}$Xi'an Jiaotong University\qquad
  $^{4}$Shanghai Innovation Institute}

\maketitle
\begingroup
\renewcommand{\thefootnote}{\fnsymbol{footnote}}
\footnotetext[1]{Equal contribution. The Acknowledgments section lists the
  author contributions.}
\footnotetext[2]{Work done while at Tencent.}
\footnotetext[3]{Corresponding authors.}
\endgroup
\thispagestyle{plain}
\pagestyle{plain}

\begin{abstract}
Reinforcement learning (RL) of LLMs computes each token's probability
twice, once in rollout and once in the trainer, and expects the two to be
\emph{bitwise consistent}. In practice they differ, and this
training--inference mismatch (TIM) can make training collapse. TIM mainly
comes from parallelism. An engine splits its computation across GPUs, runs
the pieces in parallel, and combines their results by a reduction, mostly
an addition. Floating-point addition is non-associative, so adding in
different orders gives different results. Existing work takes two
directions. The first forces rollout and the trainer to add in one agreed
order. Each engine then gives up its own configuration (e.g., its batch
size and parallel degrees), which can cost up to 91\% of the training
throughput. The second keeps parallelism and mitigates TIM in the
algorithm, which adds up to 47\% to the step time and leaves the mismatch.
Both thus trade parallelism against bitwise consistency, and the two
appear to conflict.

Our key insight is that what really conflicts with bitwise consistency is
not parallelism, but the non-associative reductions in parallelism. If
these reductions are eliminated, the two can coexist. We present \sys{},
which makes these reductions associative with two techniques.
\emph{Structural elimination} changes how a computation is split, so that
each GPU computes complete outputs and the reduction is a gather without
any addition. \emph{Arithmetic elimination} changes the operation of a
reduction to integer addition, which is associative. With \sys{}, the two
engines return bitwise-identical results under their own configurations,
and \sys{} achieves the performance of the unmodified engines
(0.98--1.04$\times$ their step time). In 300 RL steps on Qwen3-30B-A3B,
training on the unmodified engines collapses, with and without a standard
algorithmic mitigation, and training with \sys{} keeps improving.
\end{abstract}

\section{Introduction}
\label{sec:intro}

\begin{figure*}[t]
  \centering
  \includegraphics[width=\textwidth]{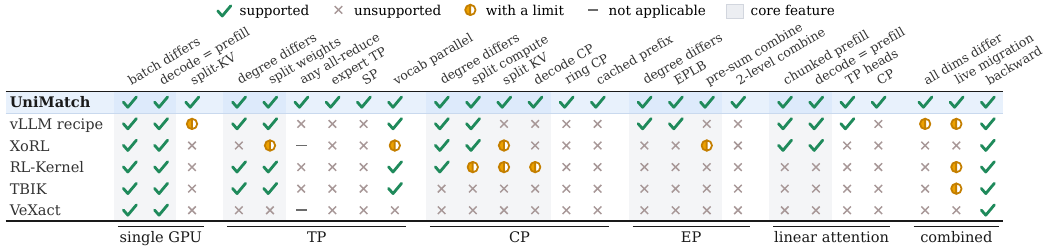}
  \caption{Features with which each system keeps training and rollout
  bitwise identical. A check means supported, a half disc supported with a limit, a cross unsupported, and a dash not applicable.}
  \label{fig:features}
\end{figure*}

Reinforcement learning (RL) has become a standard stage of LLM
post-training~\cite{ouyang2022instructgpt,shao2024deepseekmath,deepseekr1}.
RL computes each response token's probability twice. Rollout uses it to
sample the token, and the trainer computes it again for the gradient.
With the same weights, RL algorithms expect both computations to return
the same probability, i.e., rollout and the trainer should be
\emph{bitwise consistent}. Otherwise, the difference changes clipping
decisions, sampled tokens, and MoE routing~\cite{vexact2026,r3moe}, and its
errors compound along a sequence and across updates. Training can then
degrade or collapse~\cite{vexact2026,beyondprecision}.

Unfortunately, in practice, rollout and the trainer usually return
different probabilities even with the same weights and tokens. This
difference is called the \emph{training--inference mismatch} (TIM). Our
analysis shows that TIM mainly comes from parallelism
(\autoref{sec:moti-cause}). To run fast, an engine (of rollout or the
trainer) splits its computation across GPUs and across the cores of each
GPU, runs the pieces in parallel, and combines their results (i.e., a
\emph{reduction}).\footnote{The collective communication primitives
all-reduce and all-gather are both reductions in this general sense.}
Most reductions combine the results by addition. However, floating-point
addition is non-associative due to the limited precision of hardware, so
adding in different orders gives different results, which causes TIM.

Existing work handles TIM in two directions. The first direction trades
parallelism for bitwise consistency. These methods force rollout and the
trainer to add in one agreed
order~\cite{vexact2026,tmlbio,deepseekv4,vllmbitwise2026,tbik2025,xorl2026}
(e.g., both compute $(a+b)+c$, and neither computes $a+(b+c)$), so that
both return the same result for every reduction, which avoids TIM.
However, rollout and the trainer have different workloads. Rollout
generates one token at a time, and the trainer runs forward and backward
over whole sequences. They therefore usually use different engines (e.g.,
vLLM for rollout and Megatron-Core for the trainer). To get the best
performance, each engine runs with its own \emph{configuration} (e.g., its
batch size and parallel degrees), which determines the order of its
reductions. Thus, when forced to follow an agreed order, an engine has to
use a suboptimal configuration and give up the run-time optimizations that
conflict with this order. This can cost up to 91\% of the training
throughput~\cite{vexact2026} (\autoref{sec:eval-perf}). Furthermore, in
some scenarios, an agreed order cannot be achieved at all. For example,
in context parallelism (CP), training splits a long sequence by its final
length, which rollout does not know while it generates the sequence
(\autoref{sec:moti-cost}, \autoref{fig:features}).

The second direction is algorithmic mitigations, which keep parallelism
and try to fix the inconsistency. These methods let rollout and the
trainer return different results, and they reduce the mismatch or correct
the update for
it~\cite{tis2025,beyondprecision,fp16rl,trm,mipi,r3moe,minimaxm1,deepseekv32}
(e.g., by weighting each token by the clipped ratio of its two
probabilities~\cite{tis2025}).
In practice, performance, which mainly comes from parallelism, is critical for
training, so most production RL systems take this
direction~\cite{modal2026,scorecentering2026}.
However, as our measurements show, these mitigations still add up to 47\% to
the step time, and even so the mismatch remains and training can still
collapse (\autoref{sec:moti-mitigation}).

Both directions thus trade parallelism against bitwise consistency, and the
two appear to conflict.
Our key insight is that what really conflicts with bitwise consistency is
not parallelism, but the non-associative reductions in parallelism. If we
can eliminate these reductions, parallelism and bitwise consistency can
coexist. Specifically, parallelism cannot avoid reductions, because the
pieces must be combined. However, the result of a reduction depends on the
order when its operation is non-associative (e.g., floating-point
addition). We call such a reduction a \emph{non-associative reduction}.
With an associative operation, a reduction gives the same result in every
order. Therefore, the fundamental way to solve TIM is to identify the
non-associative reductions and make them associative. An RL system that
eliminates them in this way keeps its parallelism and is bitwise
consistent.

We present \sys{}, an RL system built on this insight.
\sys{} makes rollout and training bitwise identical
under each engine's own configuration and keeps the performance and accuracy
of the unmodified engines.
Based on two further observations (\autoref{sec:insight}), \sys{} uses two
techniques to eliminate non-associative reductions.

\emph{Structural elimination} changes how a computation is split. Some
computations can be split so that each GPU computes complete outputs
(e.g., whole columns of a matrix product), with the same load on each GPU.
The engine then combines the outputs by a gather without any addition, and
a gather is associative.
However, the unmodified engines split the sum on purpose to gain performance. A
split by outputs can leave too little parallelism, needs more
communication, and changes the layout that other optimizations rely on.
\sys{} reduces these costs by overlapping the communication with computation
and by several other optimizations (\autoref{sec:design-tp}).

\emph{Arithmetic elimination} changes the operation of a reduction. When a
computation cannot be split by its outputs, \sys{} converts the values to
integers before combining them, because integer addition is associative.
We observe that, at the right scale, integers of moderate width keep the
precision of the floating-point reduction. However, LLM engines are built
around floating point, both engines must choose the scale before all
values exist, and integers take more bits than the native format. \sys{}
uses integers only where partial results are combined, and computes the
scale on each GPU from values that it already holds
(\autoref{sec:design-cp}, \autoref{sec:design-ep}).

We evaluate \sys{} on Qwen3-30B-A3B and Qwen3.6-35B-A3B, with vLLM for
rollout and Megatron-Core for training. First, the two engines return
bitwise-identical results under different TP, CP, and EP degrees, with
their run-time optimizations on (\autoref{fig:features}). Second, \sys{}
achieves the performance of the unmodified engines (0.98--1.04$\times$
their step time across models and workloads). Third, in 300 steps of RL on
Qwen3-30B-A3B, \sys{} keeps improving, and training on the unmodified
engines collapses with and without truncated importance sampling
(TIS)~\cite{tis2025} (\autoref{fig:longrun}).

To the best of our knowledge, \sys{} is the first RL system that achieves
the following three at once: (1)~it is bitwise consistent between rollout
and training, (2)~it lets each engine keep its own configuration, and
(3)~it achieves the performance of the unmodified engines.

In summary, this paper makes the following contributions:
\begin{itemize}
  \item A characterization of TIM in RL, which shows that TIM comes mainly
  from parallelism and that existing mitigations leave it in place
  (\autoref{sec:motivation}).
  \item An analysis that reveals the importance of non-associative
  reductions in TIM, and two techniques that eliminate them, structural
  elimination and arithmetic elimination (\autoref{sec:insight}).
  \item \sys{}, a bitwise-consistent RL system that applies the two
  techniques to TP, CP, EP, and linear attention (\autoref{sec:design},
  \autoref{sec:impl}).
  \item An evaluation on two 30B-scale models, which shows that \sys{} is
  bitwise consistent under each engine's own configuration and achieves the
  performance of the unmodified engines (\autoref{sec:eval}).
\end{itemize}

\section{LLM Parallelism And TIM}
\label{sec:bg}

\subsection{Parallel Execution of LLMs}
\label{sec:bg-parallel}

An engine runs an LLM in parallel across GPUs and across the cores of each
GPU. In several forms of this parallelism, the reduction
(\autoref{sec:intro}) adds partial sums in floating point.

\para{Tensor parallelism (TP).}
TP splits each weight matrix $W\in\mathbb{R}^{K\times N}$ of $Y=XW$ across
$p$ GPUs~\cite{shoeybi2019megatron}. \emph{Column parallelism} splits $W$
by its $N$ columns, so each GPU computes $N/p$ complete columns of
$Y$. \emph{Row parallelism} splits $W$ by its $K$ input rows, so each GPU
computes a partial sum of every output, and an all-reduce adds the $p$
partial sums. TP usually runs a column-parallel projection followed by a
row-parallel one, a layout we call Col+Row, so the output shard of the
first is the input shard of the second and the pair communicates once.
\emph{Sequence parallelism} (SP)~\cite{korthikanti2023sp} splits the token
rows outside the TP region, and the all-reduce becomes a reduce-scatter.

\para{Context parallelism (CP).}
CP splits a long sequence across GPUs. The most common form of CP is ring
attention~\cite{liu2023ring}. Each GPU keeps the queries of its tokens,
passes key--value (KV) chunks around a ring, and merges the attention over
each chunk into a running output with online softmax.
Ulysses~\cite{jacobs2023ulysses} re-partitions the sequence by attention
heads before attention.

\para{Expert parallelism (EP).}
An MoE layer sends each token to its top-$k$ experts and adds their
outputs, weighted by the routing weights. EP places the experts on
different GPUs. A \emph{dispatch} sends each token to the GPUs of its
experts, and a \emph{combine} returns the weighted outputs. To reduce
combine traffic, high-performance EP libraries first sum local expert outputs, then add the
per-GPU sums across GPUs~\cite{deepep2025,vllmepdocs}.

\para{Splits inside a GPU.}
A GPU kernel runs one thread block per output tile. 
With too few tiles to fill the GPU, split-K~\cite{cutlass} lets several thread blocks
compute partial sums of the same GEMM tile over $K$, and
split-KV~\cite{flashdecoding2023} splits the sum over keys in decode
attention, where one query attends to a long KV cache. 
Runtimes choose these splits from the input shape, so they can differ
across batches.

\subsection{RL Post-Training and TIM}
\label{sec:bg-tim}

RL post-training alternates between rollout, which samples responses from
the current policy, and training, which computes gradients from
them~\cite{sheng2024hybridflow}. PPO and
GRPO~\cite{schulman2017ppo,shao2024deepseekmath} weight each token by the
importance ratio
$\rho_t=\pi^{\mathrm{train}}_\theta(a_t\mid s_t)/\mu^{\mathrm{rollout}}_{\theta_{\mathrm{old}}}(a_t\mid s_t)$,
where the denominator is the probability used to sample token $a_t$. The \emph{training--inference mismatch} (TIM) is
$\delta_t=\log\pi^{\mathrm{train}}_{\theta_{\mathrm{old}}}(a_t\mid s_t)-\log\mu^{\mathrm{rollout}}_{\theta_{\mathrm{old}}}(a_t\mid s_t)$.
Both engines run the same weights on the same tokens, so before the first
update $\rho_t=e^{\delta_t}$ should be exactly one.

TIM enters the update in two ways. First, it is part of the importance
ratio. After an update, the ratio is the change of the policy multiplied by
$e^{\delta_t}$, and this factor multiplies to $e^{\sum_t\delta_t}$ over a
response. Asynchronous RL~\cite{areal2025} changes only the first factor,
so TIM harms it as well~\cite{reinforcezero2026}. Second, the trainer
computes the gradient on tokens that rollout sampled from another
distribution. This holds also when the loss has no importance ratio, and
when the trainer recomputes the rollout probabilities itself.

\para{Existing approaches.}
Existing work handles TIM in two directions (\autoref{sec:related}).
\emph{Bitwise-consistent systems} change the engines so that $\delta_t=0$.
They force both engines to add in one agreed
order~\cite{vexact2026,tmlbio,deepseekv4,vllmbitwise2026,tbik2025,xorl2026},
for example by sharing the forward kernels~\cite{vexact2026} or by fixing
one reduction tree for all parallel degrees~\cite{tbik2025,xorl2026}.
\emph{Algorithmic mitigations} accept TIM and reduce the mismatch or
correct the update for it. Truncated importance sampling
(TIS)~\cite{tis2025} weights each token by a clipped $e^{\delta_t}$, and
routing replay (R3)~\cite{r3moe} makes the trainer use the experts that
rollout selected. Score centering (SC)~\cite{scorecentering2026} adds a
correction term to the update. Other methods mask tokens with a large
mismatch or lower the learning rate~\cite{trm,beyondprecision}.
\section{Motivation: Parallelism versus Consistency}
\label{sec:motivation}

\subsection{Where TIM Comes From}
\label{sec:moti-cause}
\label{sec:moti-plans}

\begin{figure}[!ht]
  \centering
  \includegraphics[width=\columnwidth]{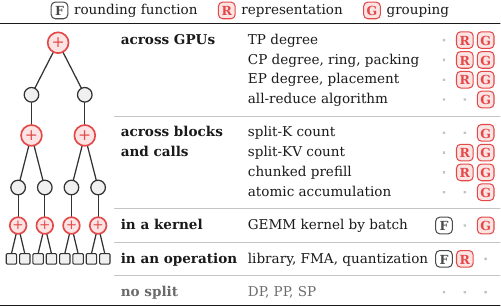}
  \caption{The settings that split a sum at each level and the kinds of
differences each one causes (F, R, G).}
  \label{fig:rootcause}
\end{figure}

We first ask why the two engines return different results. We find two
sources that cause three kinds of differences (\autoref{fig:rootcause}).

\para{The implementation of each operation.}
Two engines can compute the same operation with different code or
hardware. In software, different math libraries guarantee different
accuracy for the same function~\cite{cudamathaccuracy}, so they can return
different bits for the same input.
In hardware, Tensor Cores and CUDA cores can round at different points
when computing the same dot product, so their results can differ~\cite{tamingbitwise2026}.
This source is relatively easy to
remove, because the engines can share one implementation of each operation
with little limit on performance or parallelism (\autoref{sec:design-e2e}).

\para{Parallelism.}
An engine splits a sum into parts and adds the partial sums, across
GPUs, across the thread blocks of a GPU, and inside a kernel
(\autoref{sec:bg-parallel}, \autoref{fig:rootcause}). A split changes the
result in two ways. The first is the \emph{representation} of the partial
sums, that is, the format and scale in which they are stored. For example,
with EP, a GPU usually add the outputs of its own experts in FP32 and sends the
partial sum in BF16. The EP degree decides which experts share a GPU, so it
decides which sums are rounded to BF16, and this causes TIM. The second is
the \emph{grouping}, that is, which terms each partial sum contains and in
which order the partial sums are added (e.g., the block count of split-K).
Floating-point addition is not associative, so a different grouping also
changes the result.

The two engines could avoid this source by using the same splits. However,
they have different workloads, so each runs with its own configuration for
the best performance.
Rollout generates one token at a time for a changing batch, and training
runs forward and backward over whole sequences, so they use different
settings and optimizations. Public recipes give the two engines different
TP or EP degrees~\cite{verlrecipes,slimerecipes}, and training often uses
CP for long sequences while rollout does not. Rollout often uses split-KV in
decode and chunked prefill for long prompts, and training usually has neither. The
representation and the grouping of partial sums are thus what an approach
to TIM must handle.

\subsection{How TIM Harms RL}
\label{sec:moti-char}

We next measure how large TIM is and how it harms RL. We train
Qwen3-30B-A3B-Base on DAPO-Math-17k~\cite{dapo2025} with on-policy
REINFORCE++~\cite{reinforcepp2025} for 300 RL steps in four runs: native
(the unmodified engines), native with token-level TIS~\cite{tis2025},
native with several algorithmic mitigations combined (+allfix, i.e., R3,
TIS, ratio masking, and SC)~\cite{r3moe,tis2025,scorecentering2026}, and
\sys{}, our bitwise-consistent system (\autoref{sec:design}). To match
production recipes (\autoref{sec:moti-cause}), rollout runs vLLM at TP
degree 4, and training runs Megatron at TP, CP, and EP degrees 2, 2, and 8
(\autoref{sec:eval-setup}).

Both engines record each response token's log-probability ($\log p$). We
measure TIM by $\delta_t$, the trainer's $\log p$ of the token $t$ minus the
rollout's (\autoref{sec:bg-tim}), and report $|\delta_t|$ and $|\sum\delta|$
over a response. To compare any two runs, we replay the same responses on
native checkpoints from step 100, during normal training, and step 225, at
the onset of collapse. We compute $|\Delta\log p|$, the difference between
the two runs' $\log p$ of each token, and report its mean as a percentage
of the \emph{native difference}, the mean $|\Delta\log p|$ between native
training and native rollout at that checkpoint.

\begin{figure*}[t]
  \centering
  \includegraphics[width=\textwidth]{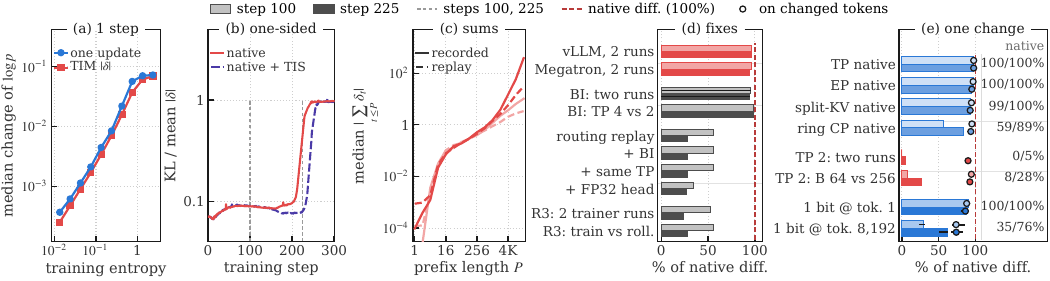}
  \caption{TIM in RL. (a)~Median change of a token's $\log p$ from one
  optimizer step and from TIM, by entropy. (b)~The one-sided part of the
  mismatch (KL divergence from rollout to training) relative to mean
  $|\delta|$. (c)~Median $|\sum\delta|$ over a response's first $P$
  tokens, recorded in training and between replays. (d)~Mean
  $|\Delta\log p|$ between two runs of one engine, and between training and
  rollout as fixes are added. BI is batch-invariant kernels, and R3 is
  routing replay. (e)~Mean $|\Delta\log p|$ after one change, which is one
  native merge in \sys{}, a second vLLM run, or one changed bit. Dots cover
  only the tokens after the first differing one, and the right column gives
  the share of the native difference on them.}
  \label{fig:mismatch-char}
\end{figure*}

\para{Finding 1: Per token, TIM is of the same order as the change of one
optimizer step.}
We compare the native $|\delta|$ with the change of a token's $\log p$ from
one optimizer step, which we measure in \sys{} because it has no TIM. Both
grow with the entropy of the token, so we compare them at equal entropy
(\autoref{fig:mismatch-char}(a)). At every entropy, the median $|\delta|$
is 65--96\% of the median change from one optimizer step. The importance
ratio (\autoref{sec:bg-tim}) therefore contains a TIM factor of the same
order as the policy change that it measures, so TIM cannot be ignored.

\para{Finding 2: TIM adds up along steps and responses.}
Rollout samples each token with its own probabilities, so it more often
samples the tokens that it rates higher than the trainer does. On the
sampled tokens, $\delta_t$ is therefore negative on average, with expected
value $-\mathrm{KL}(\mu^{\mathrm{rollout}}\,\|\,\pi^{\mathrm{train}})$.
This one-sided part is small on one token, 7--10\% of the mean $|\delta|$
during normal training (\autoref{fig:mismatch-char}(b)). However, it has
the same sign on every token, so it adds up along a response. In a replay,
neither engine samples, so the differences have no common sign and partly
cancel. Over long responses, the median recorded $|\sum\delta|$ is
3.0$\times$ that of a replay (\autoref{fig:mismatch-char}(c)). The
one-sided part also adds up along steps. At the onset of collapse, it
reaches 51\% of the mean $|\delta|$, and the recorded $|\sum\delta|$ is
12.5$\times$ that of a replay. The trainer thus updates on responses that
its own policy would rarely sample.

\para{Finding 3: With TIM, native training collapses.}
The native run peaks near step 220 and then collapses
(\autoref{fig:longrun}). Its accuracy on AIME 2024 falls from 34.3\% to
2.2\% by step 300. With the same recipe and without TIM, training keeps
improving for all 300 steps (\autoref{sec:eval-rl}).

\subsection{Why Bitwise Consistency Is Hard}
\label{sec:moti-cost}

Naturally, existing bitwise-consistent systems make both engines agree on
the representation and the grouping of every partial sum
(\autoref{sec:related}). Both engines then add in one agreed order, so TIM
does not exist within the supported configurations
(\autoref{fig:features}). For EP, vLLM's DeepSeek-V4 and Kimi K3.1 recipes
return each expert output to its token's GPU and add the outputs in a fixed
slot order~\cite{vllmbitwise2026,vllmzeromismatch2026}, so a remote GPU
cannot first add the outputs of its own experts. As MoE models use more and
larger experts, the combine sends more data, so DeepEP V2 adds a token's
expert outputs on each GPU and again within each node before sending them to
other nodes~\cite{deepepv2}, which a fixed slot order rules out. For TP,
TBIK, IsoExec, RL-Kernel, and vLLM's Kimi K3.1 recipe add in one balanced
binary tree~\cite{tbik2025,isoexec2026,rlkernel2026,vllmzeromismatch2026}, so the shard
of each GPU must be a subtree, and the communication library cannot use its
optimized all-reduce to add the shards. Inside a GPU, batch-invariant
kernels fix the splits~\cite{tmlbio,sglangdet}, so a kernel cannot choose
them from the input shape. Neither engine can choose its own splits, and an
agreed order therefore has three costs.

\para{Finding 4: An agreed order costs up to 91\% of the training
throughput.}
To add in the agreed order, an engine must turn off the optimizations that
choose splits at run time, such as split-K, split-KV, and chunked prefill.
Turning off split-KV alone costs 15--41\% when prompt lengths
vary~\cite{corun2026}. On our hardware, the forward operators of five such
systems fall to 3--19\% of native throughput on Qwen3-30B-A3B in the worst
case~\cite{vllmbitwise2026,xorl2026,rlkernel2026,tbik2025,vexact2026}, and
those of TBIK and VeXact reach at most 36\% (\autoref{fig:perf-ops}). End to
end, with the unsupported parallelism also turned off, an agreed order can
cost up to 91\% of the training throughput~\cite{vexact2026}
(\autoref{sec:eval-perf}).
With such an unacceptable performance loss, these systems are mainly used
to debug failing runs instead of in production~\cite{gdnasyncrl2026,sglangdet}.

\para{Finding 5: An agreed order also lowers accuracy.}
To agree on the representation, an engine must convert partial sums to a
format that its own splits do not need. The engine with the higher
precision follows the one with the lower precision, so it loses accuracy.
For example, a trainer at EP degree 8 that follows a rollout at EP degree
32 rounds 32 partial sums to BF16. On Qwen3-30B-A3B, BF16 rounding at each
GPU raises the mean combine error to 1.25--1.44$\times$ that of rounding
once at EP degrees 2--8 (\autoref{sec:eval-accuracy}).

\para{Finding 6: Some parallelism cannot follow an agreed order at all.}
Some splits depend on information that the other engine does not have, so
the two engines cannot agree on an order in advance. 
For example, ring CP in training
splits a sequence by its final length, which rollout does not know while it
generates the sequence. None of the existing systems in \autoref{fig:features}
supports ring CP, and the only one that runs CP in decode gathers the whole
KV cache on every GPU at each step~\cite{rlkernel2026}.

Because of these costs, most production RL systems keep their parallelism,
accept TIM, and rely on algorithmic mitigations. We next measure what this
choice costs.

\subsection{Why Algorithmic Mitigations Fall Short}
\label{sec:moti-mitigation}

Existing mitigations either correct the update or reduce the mismatch (e.g. )
(\autoref{sec:bg-tim}, \autoref{fig:e2e-systems}). We first show why the
mismatch is hard to reduce (Findings~7 and~8), and we then measure the
mitigations (Finding~9).

\para{Finding 7: A single setting difference causes a large mismatch.}
Changing one setting of a native engine, such as its CP or EP degree, batch
size, prefill chunk size, or sequence packing order, gives 95--99\% of the
native difference (\autoref{fig:mismatch-char}(d)). Even two runs of one
engine with the same configuration differ by 95--97\%.
Even with batch-invariant (BI) kernels, which remove these run-time splits,
changing only the TP degree from 4 to 2 still gives 98.3\%.

The same holds in the other direction. We start from \sys{}, which has no
mismatch, and put back the native merge of one reduction
(\autoref{fig:mismatch-char}(e)). With the native TP all-reduce, EP
combine, or split-KV alone, the mismatch returns to 93--98\% of the native
difference. Ring CP changes only the tokens after its first merge across
GPUs, and on these tokens the mismatch returns to 94--95\%.
Therefore, an approach that leaves
one difference between the engines leaves the mismatch nearly unchanged.

\para{Finding 8: A smallest difference causes a large mismatch on the
next tokens.}
A difference does not stay as small as it starts. In \sys{}, two replays
are bitwise identical. We change one element of the layer-0 input of one
token by one rounding unit in one replay. On the tokens after it, the two
replays then differ by 69--91\% of the native difference
(\autoref{fig:mismatch-char}(e)), because the difference reaches all later
tokens through attention and MoE routing. Keeping the two engines within a
small tolerance is therefore not enough.

\para{Finding 9: Existing mitigations add a significant performance cost,
and the mismatch remains.}
TIS corrects the update.
It weights each token by $\min(e^{\delta_t},2)$, which needs another
trainer forward pass and adds 15--21\% to the step time
(\autoref{sec:eval-perf}). TIS does not change the tokens that rollout
samples, so the one-sided part of Finding~2 remains. Training with TIS
collapses only 20 steps later than native training
(\autoref{fig:longrun}).

R3 reduces more of the mismatch than TIS, but it is also slower.
We combine it with TIS, ratio
masking, and score centering~\cite{scorecentering2026}, and we call the
combination +allfix. +allfix adds 35--47\% to the step time
(\autoref{sec:eval-perf}). Started at native's step 200, it avoids the
collapse (\autoref{sec:eval-rl}), but 27--56\% of the native difference
remains. Adding batch-invariant kernels~\cite{tmlbio}, the same TP degree
in rollout as in training, and an FP32 output layer~\cite{minimaxm1} still
leaves 27--34\% (\autoref{fig:mismatch-char}(d)).

Most of this remainder comes from the trainer alone. With routing replay,
two runs of the trainer on the same responses differ by 52.7\% of the
native difference, close to the 56.1\% between training and rollout.
The fixes above change the rollout side, so they cannot lower the mismatch
below the difference between two trainer runs.

Production RL systems adopt these mitigations and accept TIM to keep their
performance (\autoref{sec:moti-cost}). However, the mitigations themselves
still add 15--47\% to the step time.

\subsection{Design Goals}
\label{sec:moti-goals}

The findings above give three goals for an approach to TIM. One remaining
difference leaves the mismatch nearly unchanged (Finding~7). This gives
\textbf{goal~\step{1}}. An approach to TIM should cover every difference
between the engines. The smallest difference grows to a large mismatch on
the next tokens (Finding~8). This gives \textbf{goal~\step{2}}. The two
engines should ideally return bitwise-identical log-probabilities. RL
training occupies large GPU clusters for weeks and is sensitive to step
time~\cite{reflectionbeam2026,fp8rl,kv4rl2026,nvdetpretrain2026,megatron8261},
and both an agreed order and the mitigations add to it (Findings~4
and~9). This gives \textbf{goal~\step{3}}. An RL step should take the
native time, that is, the step time of the unmodified engines.

Neither algorithmic mitigations nor bitwise-consistent systems meet the
three goals. Algorithmic mitigations keep the parallelism of each engine,
but they miss goals~\step{1} and~\step{2}, and with their own cost they
also miss goal~\step{3}. Bitwise-consistent systems meet goals~\step{1}
and~\step{2} within their supported configurations and lose the
parallelism, so they miss goal~\step{3} (\autoref{sec:moti-cost}). We aim to
meet the three goals.
\section{Insight and Challenges}
\label{sec:insight}

As \autoref{sec:motivation} shows, both algorithmic mitigations and existing
bitwise-consistent systems trade parallelism against bitwise consistency,
and the two appear to conflict.

\para{Insight: What really conflicts with bitwise consistency is not
parallelism, but the non-associative reductions in parallelism.}
If we can eliminate these reductions, parallelism and bitwise consistency
can coexist. Parallelism must combine its partial results, so reductions
are necessary (\autoref{sec:intro}). A \emph{non-associative reduction}
uses a non-associative operation, such as floating-point addition, and its
result depends on the split. A reduction that uses an associative
operation gives the same result under every split. Eliminating a
non-associative reduction therefore means making it associative, and
there are two ways to do so (\autoref{fig:design}(a)). The first changes
the split, so that the engine combines the partial results by a gather. A
gather only concatenates them, and concatenation is associative. The
second keeps the split and changes the operation to an associative one,
such as integer addition.

Consider an output $y=a+b+c+d$. One engine computes it on one GPU as
$((a+b)+c)+d$, and another engine splits it across two GPUs as
$(a+b)+(c+d)$. The reduction of the second engine adds two partial sums in
floating point, so it is non-associative and the two results can differ.
The first way changes the split, so that one GPU holds all
four values of $y$ and the other GPUs compute other outputs. Both engines
then compute $((a+b)+c)+d$, and the engine combines the outputs by a
gather, which concatenates them without any addition.
The second way keeps the split and converts the
values to integers before adding them. Both engines then compute
$\mathrm{int}(a)+\mathrm{int}(b)+\mathrm{int}(c)+\mathrm{int}(d)$, which is
the same under every grouping.

Once every reduction is associative, every split gives the same bits, so
the engines need no agreed order. Our design principle is to \textbf{eliminate
non-associative reductions}. This covers every difference that parallelism
causes (goal~\step{1}) and makes the two engines bitwise identical
(goal~\step{2}). The challenges below concern goal~\step{3}, the native
step time. \sys{} changes the split where it can
move to an axis that indexes the outputs (\autoref{sec:chal-structural}) and
changes the operation elsewhere (\autoref{sec:chal-arith}).

\subsection{Structural Elimination}
\label{sec:chal-structural}

\para{Observation 1: Some splits can move from the reduction axis to an output
axis.}
The \emph{reduction axis} is the dimension over which values are added,
and an \emph{output axis} is an axis that indexes the outputs. Some
computations can be split along either kind of axis with the same load on
each GPU.
For example, a $K\times N$ weight matrix can be split by its $K$ rows, the
reduction axis, or by its $N$ columns, an output axis, and both splits give
each GPU the same weights and FLOPs. A split along an output axis forms no
partial sums, so the engine combines the outputs by a gather. We call this
\emph{structural elimination}. Native engines split the reduction axis for three reasons,
and each is a challenge for structural elimination.

\para{Challenge-S1: Too little parallelism on the output axis.}
An output axis can be too short to fill a GPU. Some computations have few
outputs, and each output sums over thousands of values (e.g., a decode
step has one token per sequence). A split along the output axis then
leaves most cores of a GPU idle, so native kernels split the reduction
axis (\autoref{sec:design-tp}).

\para{Challenge-S2: Extra communication for the inputs.}
A GPU that computes complete outputs needs every input of these outputs.
A split along the reduction axis lets each GPU use the inputs that it
already holds. A split along an output axis must first gather the inputs
from the other GPUs, which adds communication that the native split does
not have (\autoref{sec:design-tp}).

\para{Challenge-S3: Extra memory and a different layout.}
Each GPU must also store the gathered inputs for backward, which takes $p$
times the memory of its own share with $p$ GPUs. The split also leaves the
outputs in a different layout, and the optimizations that expect the
native layout no longer apply directly (\autoref{sec:design-tp}).

\subsection{Arithmetic Elimination}
\label{sec:chal-arith}

Some splits cannot move to an output axis. The output axis may be too
short for the parallelism that the engine needs (S1), or the engine may
split the reduction axis to divide the weights across GPUs. For these
splits, \sys{} changes the operation of the reduction. It converts the
\emph{terms} of a reduction to integers before they are combined, because
integer addition is associative. A term is a value that both engines compute identically
under any split, such as each of $a$ to $d$ in the example above. We call
this technique \emph{arithmetic elimination}.

\para{Observation 2: At the right scale, integers of moderate width keep
the precision of a floating-point reduction.}
The precision of a floating-point sum is set by its largest value, because
this value sets the exponent and the format keeps a fixed number of
significant bits below it. For example, native engines often round partial
sums to BF16, which keeps eight significant
bits~\cite{goldberg1991floating} of the largest value. A fixed-point
integer can keep the same bits and reserve $\lceil\log_2 n\rceil$ more bits
to add $n$ terms. Both engines convert each term at the same scale, and the
integers then add exactly. However, native engines add in floating point
for three reasons, and each is a challenge for arithmetic elimination.

\para{Challenge-A1: Floating-point infrastructure.}
GPU hardware, kernels, and communication libraries for LLMs are built for
floating point, so integers should appear in as little of the computation
as possible. The values must stay in floating point until the first
reduction that depends on the split, and only this reduction can use
integers. The integers must also fit the registers and the communication
path (\autoref{sec:design-cp}, \autoref{sec:design-ep}).

\para{Challenge-A2: Choosing a scale before the terms exist.}
Every GPU must use the same scale from its first conversion. At that time
it has seen only some of the terms, because the others are on other GPUs
or do not exist yet. The exact scale comes from the largest term, and
finding it needs another pass and communication. A loose upper bound
prevents overflow but loses precision (\autoref{sec:design-cp},
\autoref{sec:design-ep}).

\para{Challenge-A3: Integer width.}
A wider integer allows a looser scale and a larger sum, but it costs
registers, memory, and bandwidth. For example, a 32-bit integer doubles
the payload of a BF16 value (\autoref{sec:design-cp},
\autoref{sec:design-ep}).

\section{Design}
\label{sec:design}

\begin{figure*}[htbp]
  \centering
  \includegraphics[width=\textwidth]{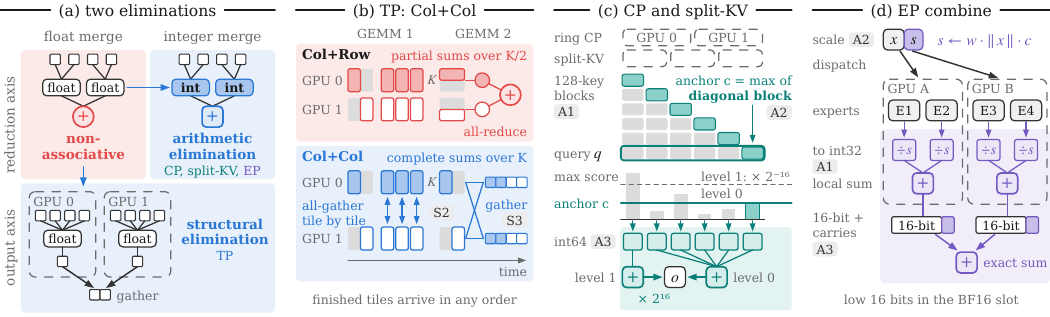}
  \caption{(a)~The two ways to eliminate a non-associative reduction.
  (b)~TP with Col+Row or Col+Col and distributed epilogues.
  (c)~Attention blocks, the anchor $c$ from the diagonal block, and levels.
  (d)~Expert combine with a per-token scale $s$ and sparse carries.
  Shading marks integer operations.}
  \label{fig:design}
\end{figure*}

\begin{figure}[htbp]
  \centering
  \includegraphics[width=\columnwidth]{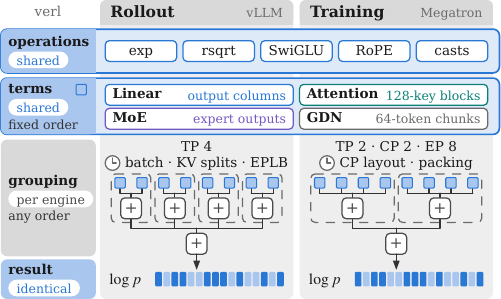}
  \caption{\sys{} in an RL system, drawn by a sum's layers. The engines share the blue rows, and each chooses the gray row, including run-time choices (clock). Term names use the colors of \autoref{fig:design}.}
  \label{fig:overview}
\end{figure}

\sys{} eliminates the non-associative reductions of each form of
parallelism, and each engine keeps its own configuration
(\autoref{fig:overview}). It applies structural elimination to TP and
split-K (\autoref{sec:design-tp}), arithmetic elimination to CP, split-KV,
and EP (\autoref{sec:design-cp}, \autoref{sec:design-ep}), and both to linear
attention (\autoref{sec:design-gdn}). \autoref{sec:design-e2e} covers the rest
of the model.

\subsection{TP: Sending Finished Tiles}
\label{sec:design-tp}

We first address the TIM that TP causes. In native Col+Row TP, the second
projection of each FFN splits $K$, its reduction
axis (\autoref{sec:bg-parallel}, \autoref{fig:design}(b), top). Each GPU
rounds a partial sum over its $K/p$ rows to BF16, where $p$ is the TP
degree, and an all-reduce adds the partial sums. This all-reduce is a
non-associative reduction. The TP degree decides which partial sums are
rounded, so two TP degrees give different bits.

We notice that this reduction comes only from the row split. \sys{}
therefore splits both projections by columns (\emph{Col+Col},
\autoref{fig:design}(b), bottom). Each GPU computes its output columns over
the full $K$, and the engine combines the columns by a gather. No partial
sum exists, so the reduction is associative.

\para{Distributed epilogues (S2).}
However, compared with Col+Row, Col+Col needs one more gather, that of the
intermediate activation. \sys{} overlaps this transfer with the GEMM that
produces the activation. A GEMM kernel computes its output in tiles. In
Col+Row, a tile of the second projection holds only a partial sum. With
Col+Col, a tile is a complete output as soon as its loop over $K$ ends. The
\emph{distributed epilogue} of the kernel then writes the finished tile
directly into the buffers of the other GPUs while the kernel computes the
next tiles. The receivers can store the tiles in any arrival order. A
Col+Row kernel can overlap computation and transfer in the same
way~\cite{flux2024,cutlassdistgemm}.

\para{Restoring the SP layout (S3).}
Col+Col also changes the layout of the output, which the optimizations
around TP depend on. It leaves output columns on each GPU, and SP expects
token rows (\autoref{sec:chal-structural}). \sys{} restores the SP layout in
the second projection. Its distributed epilogue sends each finished tile to
the GPU that owns the tile's token rows. For backward, each GPU saves only
its own slice of the intermediate activation.

\para{Short output axes in decode (S1).}
In decode, a GPU has few token rows and $N/p$ output columns, so native
kernels split $K$ to use more thread blocks (\autoref{sec:bg-parallel}).
\sys{} keeps the full loop over $K$ in one accumulator and uses narrower
output tiles, which gives more parallelism over the columns. Prefill and
decode therefore give the same bits (\autoref{tab:consistency}).

\subsection{CP and Split-KV: A Local Anchor}
\label{sec:design-cp}

We next address the TIM that CP and split-KV cause. Attention computes each
output as a sum over the keys, and native ring CP and split-KV split the
keys, its reduction axis (\autoref{sec:bg-parallel}). Each part scales its
partial sums by its own maximum score. Online softmax then combines the
parts in floating point, and it rescales the running sums whenever a new
part has a larger maximum. This combination is a non-associative reduction.
The CP degree and the split count decide the parts and the order in which
they are combined, so two splits give different bits.

We notice that structural elimination does not apply here, because the
heads, the output axis of attention, limit the parallelism to the head
count (\autoref{sec:chal-arith}). \sys{} therefore applies arithmetic
elimination. For a query with scores $x_j$ and value vectors $v_j$,
attention returns $o=\sum_j e^{x_j-c}v_j/\sum_j e^{x_j-c}$ for any offset
$c$. \sys{} uses the same $c$ in every part and adds both sums as integers
(\autoref{fig:design}(c)).

\para{A key grid (A1).}
However, attention kernels compute in floating point, so \sys{} uses
integers only above a smallest unit. It places a \emph{grid} on absolute
key positions and takes one \emph{block} of 128 keys, the tile of
FlashAttention-3~\cite{fa3}, as this unit. Each engine can split the keys
at any block boundary, so a block holds the same keys under any split. The
kernel computes the numerator and the denominator of each block in
floating point, and both engines get the same bits. \sys{} converts these
values to fixed-point integers and adds them across blocks.

\para{A local anchor (A2).}
However, both engines must choose the same $c$ before they have seen all
the keys of a query. The exact row maximum needs another pass and, under
CP, an all-reduce. We tried two cheaper upper bounds. A norm-based
(Cauchy--Schwarz) bound is so loose that most rows round to zero
(\autoref{sec:eval-ablation}), and a low-precision pass over the scores is
still too slow. We notice that $c$ does not need to be an upper bound, and
the maximum over part of the keys is enough. \sys{} therefore uses a local
anchor, the largest visible score in the block that contains the
query itself. The row maximum is usually close to this value, and the
query's GPU already holds this block at the first ring step. The anchor is
one of the row's scores, so at least one exponential is one and the
denominator does not round to zero. It uses only the current and earlier
keys, so decode and training compute the same anchor.

\para{Levels for large terms (A2).}
The anchor can be far below the row maximum, and the exponentials then
exceed the integer range. On Qwen3-30B-A3B, the row maximum exceeds the
anchor by 11.4 bits at the 99th percentile. \sys{} gives each block a
\emph{level} $\ell$ and scales its exponentials by the exact power of two
$2^{-16\ell}$. The level depends only on the block's scores and the row's
anchor, so both engines assign the same level under any split. The integers
of each level are added separately and combined at the end with exact
powers of two. Over 99\% of the rows use only level~0
(\autoref{sec:eval-ablation}).

\para{Integer width (A3).}
\sys{} uses 64-bit integers. A 32-bit integer cannot hold a sum over a long
sequence, and 64 bits is the widest integer that a GPU adds natively. With
a power-of-two scale on the value vectors, neither sum can overflow for any
input. The cost is on-chip resources, because 64-bit accumulators take more
registers and shared memory than the floating-point ones. \sys{} therefore
keeps only level~0 in registers and writes the rare higher levels to GPU
memory.

Attention therefore gives the same bits under any CP degree and split
count, in prefill and in decode, and each engine chooses them at run time
(\autoref{tab:consistency}).

\subsection{EP: Local Sums with Sparse Carries}
\label{sec:design-ep}

We then address the TIM that EP causes. Native EP adds a token's expert
outputs on each GPU, rounds the sum to BF16, and then adds the sums across
GPUs (\autoref{sec:bg-parallel}, \autoref{fig:design}(d)). This addition is a
non-associative reduction. The EP degree and the expert placement decide
which outputs are rounded together, so two placements give different bits.

We notice that structural elimination does not apply here too, because EP
splits the experts to divide their weights (\autoref{sec:chal-arith}).
\sys{} therefore applies arithmetic elimination. It keeps the local
addition, which avoids sending every expert output to its token's GPU
(\autoref{sec:moti-cost}), and adds a token's expert outputs as integers on
each GPU and across GPUs.

\para{Integers after the expert GEMM (A1).}
However, expert kernels compute in floating point, so \sys{} uses integers
only above a smallest unit. This unit is one weighted expert output, which
is the same under any EP degree and placement. \sys{} converts it to
integers right after the second GEMM, before any local addition. We call
the additions that follow the \emph{integer combine}.

\para{A scale before dispatch (A2).}
However, all GPUs must convert a token's outputs at the same scale $s$, and
a GPU cannot see remote expert outputs before it converts its own. \sys{}
therefore computes $s$ from an upper bound of the outputs, which the
token's GPU knows before dispatch and sends with the token. The bound uses
only the routing weight, the norm of the token, and a constant of each
expert that depends only on its weights. This bound is loose and costs
precision, so \sys{} adds in 32 bits. A term then keeps at least eight
significant bits when the bound is below $2^{20}$ times the largest
output, and the layer medians on Qwen3-30B-A3B are $2^{13.6}$ to
$2^{16.4}$.

\para{Sparse carries over a 16-bit payload (A1, A3).}
However, sending 32-bit sums would double the native BF16 payload. We
notice that most channel sums fit in 16 bits, because large activations
concentrate in few
channels~\cite{dettmers2022llmint8,sun2024massive}. Each GPU therefore
sends only the low 16 bits of each channel sum. If a sum does not fit, the
GPU also sends the channel index and the high bits, which we call a
\emph{carry}. Carries are rare (\autoref{sec:eval-ablation}). The low bits
fit a native BF16 payload slot, and the receiver adds them exactly in
FP32~\cite{deepep2025}.

The combine therefore gives the same bits under any EP degree, any expert
placement, including EPLB~\cite{eplb2025}, and any order of the additions
on a GPU (\autoref{tab:consistency}).

\subsection{Linear Attention: A Chunk Grid}
\label{sec:design-gdn}

We finally address the TIM that linear attention causes, with Gated
DeltaNet (GDN)~\cite{gdn2025} as the example. A GDN layer updates a state
with one matrix product per token, so the state is a reduction over the
tokens by multiplication. Native prefill computes this reduction in
64-token chunks, and decode computes it one token at a
time~\cite{flashinfer2025,fla2024}. This reduction is non-associative. The
prefill boundary decides where the state is rounded, so two boundaries
give different bits.

We notice that the two engines differ only in where they end a chunk.
\sys{} therefore places a chunk grid on absolute token
positions~\cite{xorl2026}, as it places the key grid for attention
(\autoref{sec:design-cp}). Both engines commit the state only at multiples of
64 tokens, so the state does not depend on the prefill boundary.

\para{Rows inside an incomplete chunk.}
However, decode must return the rows of a chunk before the chunk is
complete. \sys{} applies both eliminations inside a chunk. To compute row
$i$, both engines read the chunk's initial state and the cached rows
$j<i$, and then sum over all $j\le i$, so the sum is not split along its
reduction axis (structural elimination). They add the cumulative log-decay
as 64-bit fixed-point integers, so prefill and decode can group its terms
differently (arithmetic elimination).

TP splits the heads, an output axis, and under CP two head-wise all-to-all
exchanges give each GPU the whole sequences of its
heads~\cite{jacobs2023ulysses}. Linear attention therefore gives the same
bits under chunked prefill, prefix-cache hits, and decode
(\autoref{tab:consistency}).

\subsection{The Rest of the Model}
\label{sec:design-e2e}

The rest of the model reuses the techniques above. Both engines share the
implementation of each element-wise operation, such as \texttt{exp},
\texttt{rsqrt}, SwiGLU, and RoPE, which removes the first source of TIM
(\autoref{sec:moti-cause}). The vocabulary-parallel log-softmax adds its
exponentials across GPUs as integers, with the exact row maximum as the
offset. Reductions inside a GPU, such as RMSNorm, use batch-invariant
kernels~\cite{tmlbio}. Column-split expert GEMMs under TP give the same
expert outputs as EP. Col+Col also applies to the attention output
projection. This projection is a single small GEMM with little computation
to overlap its transfers with (S2), so \sys{} also supports a row split
that adds eight segments of $K$ in one tree~\cite{tbik2025}, which gives
the same bits as Col+Col at TP degrees 2, 4, and 8.

\para{Backward pass.}
TIM concerns the two forward passes, and only the trainer runs backward, so
the bitwise guarantee covers the forward log-probabilities. In backward,
\sys{} adds the input gradients of Col+Col FFNs and of experts as 32-bit
integers, so they do not depend on the TP degree or the expert placement.
It adds router gradients over fixed batch segments in a fixed order.
Attention keeps the FlashAttention-3 backward kernel, so its gradients
depend on the CP configuration, as they do in native training.

\para{Rules shared by the two engines.}
\sys{} shares rules between the two engines only at the level of single
terms (the blue rows of \autoref{fig:overview}). These rules are the
implementation of each operation, the formulas for anchors and scales, and
the grid positions. None of them depends on batch boundaries or parallel
degrees, so both engines compute each term identically, and each groups
the terms freely (the gray row). Existing bitwise-consistent systems must
also share the gray row, that is, the representation and the grouping of
partial sums (\autoref{sec:moti-cost}).

With all of the above, \sys{} is bitwise consistent end to end. Rollout and
training return the same bits for every token's log-probability, under each
engine's own TP, CP, and EP degrees and run-time optimizations
(\autoref{tab:consistency}).
\section{Implementation}
\label{sec:impl}

We implement \sys{} for NVIDIA Hopper GPUs in about 43,000 lines of Python,
CUDA, and C++. The GEMM and attention kernels are written in CuTe
DSL~\cite{cutlass}, and the integer combine builds on
DeepEP~\cite{deepep2025}. \sys{} uses vLLM~\cite{kwon2023vllm} as the
rollout engine and Megatron-Core~\cite{shoeybi2019megatron} as the trainer
(\autoref{fig:overview}), with verl~\cite{verl} for the RL loop. It
replaces the linear, attention, GDN, and MoE operators through the existing
layer interfaces, so both engines keep their run-time optimizations, such
as continuous batching, chunked prefill, prefix caching, EPLB, and
speculative decoding in vLLM.

\section{Evaluation}
\label{sec:eval}

We evaluate \sys{} against the three design goals
(\autoref{sec:moti-goals}) and ask four questions. Do the two engines return
identical bits under different parallelism, as goals~\step{1} and~\step{2}
require (\autoref{sec:eval-consistency})? Does \sys{} keep the accuracy of
native runs, which an agreed order lowers (\autoref{sec:eval-accuracy})? Does
an RL step take the native time, as goal~\step{3} requires
(\autoref{sec:eval-perf})? Does removing TIM change RL training
(\autoref{sec:eval-rl})?

\begin{figure*}[t]
  \centering
  \includegraphics[width=\textwidth]{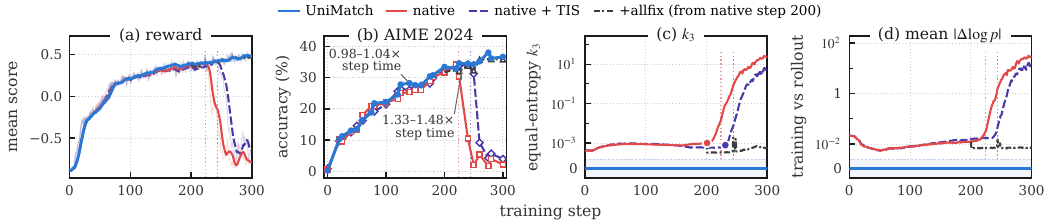}
  \caption{The 300-step runs on Qwen3-30B-A3B. (a)~Mean step score. (b)~Accuracy on AIME 2024. (c)~$k_3$ at equal entropy.
  (d)~Mean $|\Delta\log p|$ between training and rollout. Dotted lines mark
  steps 224 and 244, after which the native runs collapse. \sys{} is 0 throughout (c) and (d).}
  \label{fig:longrun}
\end{figure*}

\subsection{Setup}
\label{sec:eval-setup}

\para{Hardware and software.}
Eight servers each have eight H20 GPUs connected by NVLink, with RDMA
between servers. Operator tests use one server except at EP degrees
16--64. RL runs use all 64 GPUs.

\para{Models.}
We test Qwen3-30B-A3B~\cite{qwen3moeconfig}, an MoE model with standard
attention, and Qwen3.6-35B-A3B~\cite{qwen36config}, which also has GDN
layers. The 300-step runs and the ablations use Qwen3-30B-A3B, and dense
FFN tests use Qwen3.5-4B.

\para{Baselines.}
The native operator baselines are vLLM Col+Row and Megatron expert modules
(TP), TE ring attention on FlashAttention-3~\cite{fa3} (CP),
FlashInfer~\cite{flashinfer2025} (split-KV), and DeepEP's BF16 combine
(EP), each in its fastest configuration. The algorithmic baselines are TIS
with threshold 2~\cite{tis2025}, recomputing the old log-probabilities in
the trainer, and +allfix (\autoref{sec:moti-mitigation}), which we also time
with batch-invariant kernels (+BI). The bitwise-consistent baselines are
the vLLM DeepSeek-V4 recipe, XoRL, RL-Kernel, TBIK, and
VeXact~\cite{vllmbitwise2026,xorl2026,rlkernel2026,tbik2025,vexact2026}.
Where a system does not support our models, we port its kernels and
reduction order and verify the bitwise equalities that it claims.

\begin{figure*}[t]
  \centering
  \includegraphics[width=\textwidth]{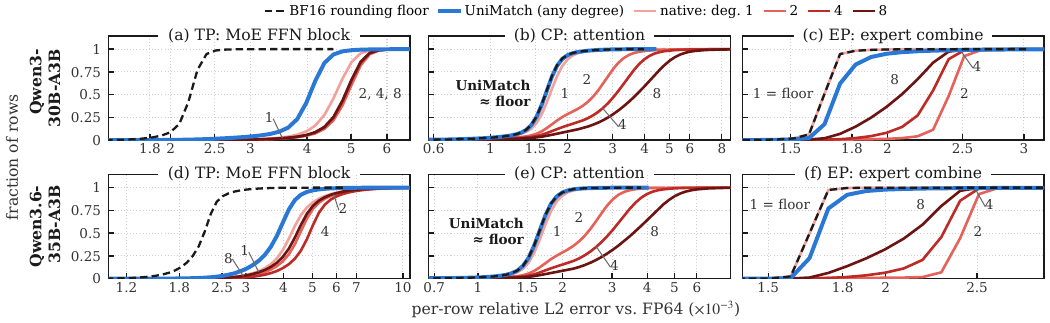}
  \caption{CDF of per-row relative L2 error against FP64 on Qwen3-30B-A3B
  (top) and Qwen3.6-35B-A3B (bottom). Columns show the TP MoE FFN block,
  attention, and the EP combine, over all layers of each model. Dashed
  curves show BF16 rounding floors.}
  \label{fig:accuracy}
\end{figure*}

\para{RL workload.}
The 300-step runs use the recipe and the parallel degrees of
\autoref{sec:moti-char}, which follow VeXact's Qwen3-30B-A3B-Base recipe on
DAPO-Math-17k~\cite{vexact2026,dapo2025}. Each of 512 prompts yields 16
responses of up to 20,480 tokens, scored $+1$ if correct and $-1$
otherwise. The loss weights each token's log-probability by its advantage
and has no importance ratio. We run \sys{}, native, and native with TIS
once each for 300 steps, and +allfix once from native's step-200 checkpoint
to step 300.

\subsection{Bitwise Consistency}
\label{sec:eval-consistency}

\sys{} returns identical bits in every pair of configurations that we
test, for single operators and for whole models.

\begin{table}[t]
  \centering
  \small
  \setlength{\tabcolsep}{3.7pt}
  \begin{tabular}{@{}llrrr@{}}
    \toprule
    & & \multicolumn{2}{c}{Identical pairs} & Native \\
    \cmidrule(lr){3-4}
    & Configurations compared & \sys{} & Control & equal (\%) \\
    \midrule
    TP & degrees 1--8 & 180/180 & 0/180 & 0--77 \\
       & prefill, decode kernels & 384/384 & & 64--100 \\
    \addlinespace
    CP & ring, chunks, split-KV & 3,844/3,844 & 0/2,964 & 51--100 \\
    \addlinespace
    EP & degrees 1--8, both engines & 72/72 & 0/9 & 58--86 \\
       & random placement & 36/36 & & 52--81 \\
       & migrated experts & 5/5 & & 2--38 \\
    \bottomrule
  \end{tabular}
  \caption{Compared pairs of configurations on Qwen3-30B-A3B. \emph{Identical
  pairs} counts the pairs with bitwise-identical outputs. Each control
  removes one mechanism of \sys{}. \emph{Native equal} is the share of equal native elements on the same pairs.}
  \label{tab:consistency}
\end{table}

\para{Operators.}
All pairs of operator configurations match on Qwen3-30B-A3B and Qwen3.6-35B-A3B
(\autoref{tab:consistency}).
The pairs cover TP, CP, and EP degrees 1--8, prefill versus decode, up to
128 KV splits, expert migration, and GDN prefix-cache hits. The TP MoE
block and the EP MoE layer match in every layer of both models. A
\emph{control} is \sys{} with one of its mechanisms removed. Every control
differs on all its pairs, so each mechanism is needed. Among the systems in
\autoref{fig:features}, only \sys{} supports ring CP, and only \sys{} runs
CP in decode without gathering the whole KV cache on each GPU.

\begin{figure*}[t]
  \centering
  \includegraphics[width=\textwidth]{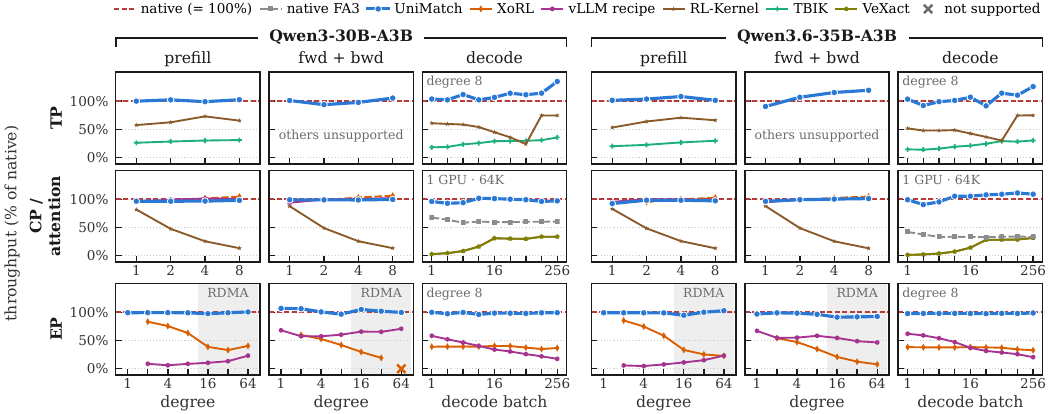}
  \caption{Throughput relative to native (100\%) for TP, attention, and EP, on Qwen3-30B-A3B (left) and
  Qwen3.6-35B-A3B (right). Columns show prefill forward and forward with backward by parallel degree, and decode by batch size.
  Crosses mark unsupported configurations.}
  \label{fig:perf-ops}
\end{figure*}

\para{Whole models.}
Across the 300 RL steps and 16.6 billion response tokens, vLLM and Megatron
return the same result for every token, at different TP, CP, and EP
degrees. Native disagrees on 88.2\% of the tokens, and native with TIS on
93.5\%. On Qwen3.6-35B-A3B, all pairs of whole-model rollout
configurations match, including EPLB, prefix caching, and multi-token
prediction, as do seven vLLM--Megatron comparisons and five RL steps.

\subsection{Accuracy}
\label{sec:eval-accuracy}

In the forward pass, the mean error of \sys{} is lower than native. Integer
conversion can give a result that differs from the native floating-point
result, so we compare both with FP64 on 8,192 MATH-500 tokens in all
layers of each model. We report the per-row relative L2 error and the share
of correctly rounded outputs (CR). For all three operators on both models,
the mean row error of \sys{} is lower than native at every degree above 1
(\autoref{fig:accuracy}). In attention, \sys{} avoids the BF16
rounding of the native ring at each step. Its CR on Qwen3-30B-A3B is
95.7\%, against 50.8\% for the native ring at CP degree 8. Given identical
expert outputs, the integer combine gives 83.5\% CR on Qwen3-30B-A3B and
87.4\% on Qwen3.6-35B-A3B, against at most 72.7\% for DeepEP's BF16 combine
at EP degrees 2--8.

\para{Tail errors.}
The loose EP bound costs precision in rare cases (\autoref{sec:design-ep}).
It leaves fewer than eight bits when it exceeds the largest expert output
by a factor above $2^{20}$, which occurs in $1.07\times10^{-4}$ of
token--layer pairs on Qwen3-30B-A3B. At EP degree 8, 0.03\% of rows
therefore exceed the native maximum error. This tail does not raise the
mean row error, which stays below native (\autoref{fig:accuracy}).

\para{GDN and backward.}
On GDN layers, the mean row error of \sys{} is
$2.16$--$3.78\times10^{-3}$, in the same range as the native prefill and
decode kernels ($2.19$--$5.44\times10^{-3}$). In backward, the MoE input
gradient at TP degree 4 has an error of $5.16\times10^{-3}$, against
$5.40\times10^{-3}$ for native. At EP degree 8, the gradient is less
accurate than Megatron's ($5.02$ versus $3.82\times10^{-3}$ on
Qwen3-30B-A3B and $6.12$ versus $4.08\times10^{-3}$ on Qwen3.6-35B-A3B).
The two errors are of the same order, and backward does not affect TIM,
which concerns the forward log-probabilities (\autoref{sec:design-e2e}).

\begin{figure*}[t]
  \centering
  \includegraphics[width=\textwidth]{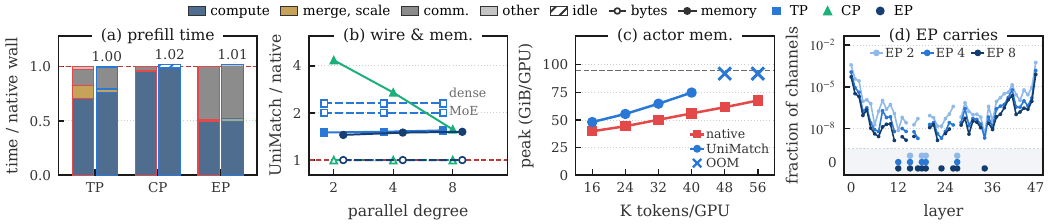}
  \caption{Costs of \sys{} on Qwen3-30B-A3B. (a)~Time of a prefill at
  degree 8 by category, relative to native wall time. Numbers give the
  wall-time ratio of \sys{}. (b)~Bytes sent and peak memory of one GPU,
  relative to native. (c)~Peak memory that the actor allocates in one long
  training step. (d)~Fraction of channels that need a carry in each layer.
  Layers that need none are at 0.}
  \label{fig:costs}
\end{figure*}

\begin{figure*}[t]
  \centering
  \includegraphics[width=\textwidth]{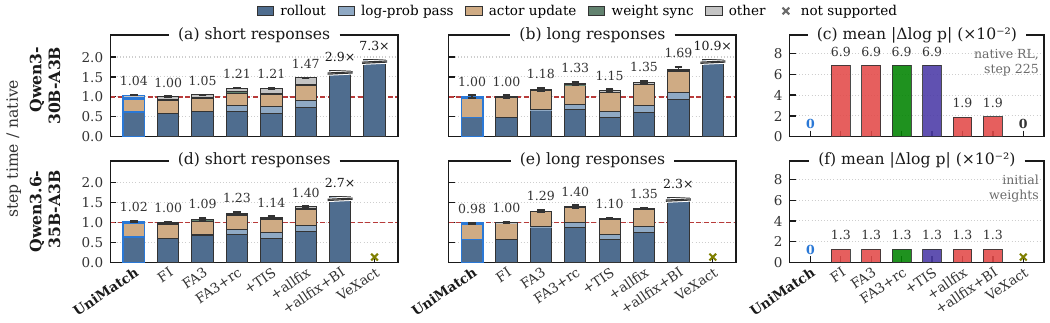}
  \caption{RL step time on 64 H20 GPUs, relative to the native step
  with FlashInfer, when every system replays the same responses. Bars stack the phases of a step, and whiskers span the measured steps.
  Rows are the two models, and the first two columns are short and long
  responses. (c) and (f)~Mean $|\Delta\log p|$ between training and rollout
  at Qwen3-30B-A3B's step-225 native checkpoint and Qwen3.6-35B-A3B's
  initial weights.}
  \label{fig:e2e-systems}
\end{figure*}

\subsection{Performance}
\label{sec:eval-perf}

An RL step of \sys{} takes 0.98--1.04$\times$ the native time on both
models.

\para{Operators.}
\sys{} reaches 93--135\% of native throughput on Qwen3-30B-A3B and
90--126\% on Qwen3.6-35B-A3B (\autoref{fig:perf-ops}). The other five
bitwise-consistent systems fall to 3--19\% of native throughput in the
worst case, a loss of up to 97\% (Finding~4, \autoref{sec:moti-cost}). On
Qwen3-30B-A3B, the forward pass of \sys{} takes 0.985--1.037$\times$ the
native time for 8,192-token prefill of the TP MoE block, ring attention,
and the EP MoE layer at degrees 1--64. The integer combine adds at most 2\%
to the MoE layer. Two cases differ. Decode at TP degree 8 takes
0.74--0.98$\times$ the native time, because one kernel gathers all output
columns. Small prefills at degrees 2--8 take 0.97--1.16$\times$, and the
whole step still takes the native time, as we show below.

\para{Other costs.}
\sys{} has other costs in communication, backward, and memory, and none of
them changes the step time. At degrees 2--8, Col+Col sends
2.00--2.30$\times$ the native bytes (\autoref{fig:costs}(b)). The
distributed epilogue overlaps these transfers with computation, so the MoE
prefill at TP degree 8 still takes 1.00$\times$ the native time. CP and EP
send the native volume. The combine and scale steps outside the compute
kernels take at most 3.1\% of native wall time (\autoref{fig:costs}(a)). TP
backward takes 1.23--1.26$\times$ the native time because of an extra sum
across GPUs, but the whole update of the trainer takes only
1.02--1.04$\times$.

\sys{} uses more peak memory, but the increase does not limit our RL runs.
In one long training step, the trainer allocates 1.21--1.34$\times$ the
native peak memory (\autoref{fig:costs}(c)), and it runs out of memory only
at a micro-batch budget four times that of our RL runs. A larger budget
also brings little, because raising the native budget from 40,960 to
57,344 tokens cuts the update time by only 0.7\%. The largest increase of a
single operator is in CP prefill, up to 4.34$\times$, where the query's GPU
holds the int64 numerator until the last ring step
(\autoref{fig:costs}(b)).

\para{End to end.}
We let every system replay the same response tokens, so all systems run the
same workload (\autoref{fig:e2e-systems}). Short and long RL steps of
\sys{} take 1.036 and 1.000$\times$ the native time on Qwen3-30B-A3B (mean
lengths 909 and 5,881 tokens), and 1.023 and 0.985$\times$ on
Qwen3.6-35B-A3B (3,422 and 14,228 tokens). With unconstrained generation,
the step time per token is 0.94--1.00$\times$ native on both models.

Both existing directions are slower. The algorithmic mitigations take
1.10--1.47$\times$ the native time across the two models, and up to
2.89$\times$ with +BI
(Finding~9, \autoref{sec:moti-mitigation}). VeXact, whose rollout uses no TP,
CP, or EP, takes 7.31--10.88$\times$, a loss of up to 91\% of the training
throughput (Finding~4). The mitigations also do not always reduce the
mismatch. On Qwen3.6-35B-A3B at its initial weights, the mean
$|\Delta\log p|$ is $1.3\times10^{-2}$ for native and for every mitigation
(\autoref{fig:e2e-systems}(f)), which is consistent with Finding~7.

\subsection{Impact on RL Training}
\label{sec:eval-rl}

Without TIM, training keeps improving for 300 steps. Both native runs
collapse, with and without TIS (\autoref{fig:longrun}).

Before the collapse, \sys{} and native train alike. Over steps 1--120,
their mean scores differ by 0.012 on average, and their response pass rates
on the same prompts differ by at most 1.1 percentage points per 20-step
window, with a standard error of about 0.12 points. From step 121 to 220,
native trails in every window, by 0.72--1.84 points
(\autoref{fig:longrun}(a)).

The native runs then collapse (Finding~3, \autoref{sec:moti-char}). In both
runs, the mismatch, measured by $k_3=e^\delta-1-\delta$, rises 12--28
steps before the gradient norm and the score change
(\autoref{fig:longrun}(c)). The mean score of native falls from $+0.399$ to
$-0.784$, and TIS delays the collapse by only 20 steps. VeXact also reports
that native rollout destabilizes training on this model and
recipe~\cite{vexact2026}. \sys{} reaches a mean score of $+0.480$ over
steps 291--300 and 36.7\% on AIME 2024 at step 300, against 2.2\% for
native and 4.0\% with TIS. +allfix, started at native's step 200, also
avoids the collapse and reaches 35.8\%, but it adds 35--47\% to the step
time and leaves 27--56\% of the native difference (Finding~9). \sys{}
removes the mismatch at native step time.

\subsection{Design Choices}
\label{sec:eval-ablation}

\para{Choice of attention anchor.}
The anchor is as accurate as the exact row maximum. With a norm-based
(Cauchy--Schwarz) upper bound as the offset, the integer sums still pass
every bitwise test on Qwen3-30B-A3B, but 62.55\% of its rows
round to zero and its mean error is 0.666. In an FP64 simulation, the exact
row maximum gives a mean error of $1.630\times10^{-3}$, and the anchor
gives $1.632\times10^{-3}$. Of 12.58 million rows, 99.57\% use only
level~0, 0.43\% reach level~1, and 2 reach level~2.

\para{Sparse carries.}
Carries are rare. On all 48 layers and 65,536 real tokens of
Qwen3-30B-A3B, at most $2.20\times10^{-5}$ of the channels need a carry at
EP degrees 2--8 (\autoref{fig:costs}(d)), so the integer combine sends
1.00$\times$ the bytes of the BF16 combine.

\section{Related Work}
\label{sec:related}

\para{Bitwise-consistent systems.}
These systems force both engines to add in one agreed order
(\autoref{sec:moti-cost}). They share the forward kernels between the two
engines~\cite{vexact2026,vllmtorchtitan2025,ascendtop,detent2026,tunixzero2026,reinforcezero2026},
fix the splits inside a
kernel~\cite{tmlbio,sglangdet,deepseekv4,flashinferrl2026}, fix one
reduction tree for all parallel
degrees~\cite{tbik2025,miles2026,isoexec2026,rlkernel2026,xorl2026,vllmzeromismatch2026}, or
gather the values on one GPU before adding them~\cite{vllmbitwise2026}.
Each system therefore supports only the splits that follow its order
(\autoref{fig:features}), which can cost up to 91\% of the training
throughput (Finding~4). \sys{} eliminates the non-associative reductions,
so the two engines need no agreed order and each keeps its own
configuration.

\para{Algorithmic mitigations.}
These methods accept TIM and reduce the mismatch or correct the update for
it (\autoref{sec:bg-tim}). They recompute the old log-probabilities or apply
importance sampling~\cite{verl,slime}, mask or reweight
tokens~\cite{trm,primerlipo2026,kpop2026,triage2026}, lower the learning
rate~\cite{beyondprecision}, change the objective or the update
rule~\cite{dcmdp,acrl,mipi,scorecentering2026,alp2026}, replay MoE
routing~\cite{r3moe,pr2moe}, or raise the precision of the
engines~\cite{fp16rl,fp8rl,qarl,hifloat4,kv4rl2026,trace2026,roundingloc2027,fp16iclr2027,quads2027}.
Several of them bias the update~\cite{baselinesc2026,s3r2026,rdc2026}. Some
reports find them sufficient on small dense
models~\cite{vexact2026,modal2026}. On our 30B MoE model, they add
15--47\% to the step time and the mismatch remains (Finding~9). \sys{}
removes the mismatch at native step time.

\para{Deterministic and exact arithmetic.}
Several systems make repeated runs of one configuration bitwise
identical~\cite{nvdetpretrain2026,mok2026,llm42,opguard2026,cublasbi2026},
but they do not cover two engines with different parallelism. Reproducible
summation and integer gradient aggregation make a sum independent of its
order~\cite{ahrens2020reproducible,exblas,repdl,cccldet2026,open1b2026,rlkernel2026,switchml2021,intsgd2022}.
Fixed-shift softmax kernels still accumulate in floating
point~\cite{flashdecodingpp2024,fixedshift2026}. \sys{} applies arithmetic
elimination inside LLM operators, where both engines must choose the scale
before the terms exist (\autoref{sec:chal-arith}).
\section{Discussion and Conclusion}
\label{sec:discussion}
\label{sec:concl}

\para{Other architectures.}
The two eliminations also apply outside \autoref{sec:design}. Arithmetic
elimination applies to the attention sums of MLA~\cite{vllmdcp2026} and to
mHC~\cite{deepseekv4}. Structural elimination applies to the DSA indexer,
which can compute all heads on each
GPU~\cite{deepseekv32,deepseekv4code}. The chunk grid applies to KDA and
Mamba-2~\cite{kimik3config}. We have not implemented mHC, KDA, or Mamba-2.

\para{Limitations.}
\sys{} guarantees bitwise equality only on GPUs that give the same bits for
each instruction~\cite{nvidiatilestability}. Its fixed-point layout is set
per model, from the context length and the value ranges. It needs more peak
memory than native, which does not limit our RL runs
(\autoref{sec:eval-perf}). A new operator with a non-associative reduction
needs a new kernel in both engines.

\para{Conclusion.}
What really conflicts with bitwise consistency in LLM RL is not
parallelism, but the non-associative reductions in parallelism. \sys{}
eliminates them, so rollout and training are bitwise identical under each
engine's own configuration, at 0.98--1.04$\times$ the native step time. In
300 steps of RL, \sys{} keeps improving and native training collapses.
\section*{Acknowledgments}

We thank Qian Qiu, Yu Shi, Lvfang Tao, Lixiang Han, Renjie Mao, and Xusheng
Chen from Tencent for their help with this project. Honglin Zhu started this
project, led the early survey, and completed the TP design
(\autoref{sec:design-tp}). Tong Xin led the research and completed the
characterization of TIM (\autoref{sec:motivation}), the designs of CP, EP,
linear attention, and the rest of the model
(\autoref{sec:design-cp}--\autoref{sec:design-e2e}), the implementation
(\autoref{sec:impl}), and the evaluation (\autoref{sec:eval}).

\bibliographystyle{plain}
\bibliography{references}

\end{document}